\documentclass{article}
\usepackage{amsmath,amssymb,graphicx,mlspconf}
\usepackage{xcolor}
\usepackage{booktabs}
\usepackage{tikz}
\usetikzlibrary{arrows.meta,positioning,calc,fit}
\pgfdeclarelayer{bg}\pgfsetlayers{bg,main}
\definecolor{blA}{HTML}{4292c6}
\definecolor{blB}{HTML}{2171b5}
\definecolor{blC}{HTML}{084594}
\definecolor{gry}{HTML}{999999}

\copyrightnotice{979-8-3195-0884-3/26/\$31.00 {\copyright}2026 IEEE}
\toappear{2026 IEEE International Workshop on Machine Learning for Signal Processing, Sep.\ 28--Oct.\ 1, 2026, Atlanta, USA}

\title{PITCH-CLASS STEERING FOR DIFFUSION-BASED MUSIC GENERATION VIA LATENT-SPACE PROBES}

\twoauthors
  {Yushi Ye\sthanks{Equal contribution.} \quad Wilson Zheng$^{*}$}
  {Carnegie Mellon University \\ \{yushiye, wilsonz\}@andrew.cmu.edu}
  {Yongyi Zang$^{*}$}
  {Independent Researcher \\ zyy0116@gmail.com}

\begin{document}
\ninept
\maketitle

\begin{abstract}
Recent work on controllable music generation has focused on autoregressive models, leaving diffusion-based systems comparatively underexplored.
We present a lightweight method for steering the pitch content of audio produced by Stable Audio Open, a latent diffusion model for music synthesis.
A small convolutional probe containing approximately 125k parameters is trained to decode frame-level pitch-class activations from the model's variational autoencoder latent space, using paired audio and MIDI data.
At inference time, the frozen probe serves as a differentiable loss function: its gradient with respect to the denoising latent is used to nudge generation toward a user-specified pitch-class sequence, requiring no retraining or architectural modification of the base model.
Across 27 evaluation trials spanning 9 text prompts and 3 target melodies, probe-guided generation increases melodic coherence by 2.4$\times$ over the unguided baseline ($p < 10^{-5}$, Wilcoxon signed-rank test), demonstrating that musically meaningful structure is both recoverable and steerable in diffusion-based music latent spaces.
\end{abstract}


\section{Introduction}
\label{sec:intro}

Text-to-music generation has advanced rapidly, with systems such as MusicGen~\cite{musicgen}, AudioLDM~\cite{liu2023audioldm}, and Stable Audio Open~\cite{evans2024open} producing audio of increasing quality and diversity.
However, text prompts alone offer only coarse control over the musical content of the output.
A user can request ``a piano piece in C major'' but cannot specify a particular melodic sequence, limiting their use in composition workflows requiring precise pitch control.

Several recent studies have begun to address this gap through inference-time intervention on model activations.
Facchiano et al.~\cite{facchiano2025activation} apply activation patching to steer binary musical attributes such as tempo and timbre in MusicGen.
Singh et al.~\cite{singh2025discovering} and Paek et al.~\cite{paek2025learning} use sparse autoencoders to discover interpretable features in autoregressive music models, while Zhao et al.~\cite{zhao2025musicrfm} introduce Recursive Feature Machines for note-level control of MusicGen activations.
Panda et al.~\cite{panda2025finegrained} use linear probe weights as steering vectors for timbre and style transfer, and Jiang et al.~\cite{jiang2025composer} steer symbolic music generation by injecting mean activation vectors for composer style.
In a complementary line of work, Koo et al.~\cite{Koo2024understanding} probe individual attention heads in music transformers to map out what musical information different components encode.

These methods target autoregressive architectures exclusively.
For diffusion-based music generation, DITTO~\cite{Novack2024Ditto} and DITTO-2~\cite{Novack2024DITTO2DD} optimize auxiliary losses at inference time but require per-attribute objective functions and do not leverage the latent-space structure of the underlying variational autoencoder (VAE).
Music ControlNet~\cite{wu2024music} adds conditioning networks for melody and dynamics control but requires architectural modification and retraining.

We take a different approach, first showing that the VAE latent space of Stable Audio Open encodes pitch-class information in a form that a small convolutional probe can decode at frame level.
We then repurpose this frozen probe as a differentiable loss function during diffusion sampling: at each denoising step within a specified window, the gradient of a binary cross-entropy loss between the probe's pitch-class predictions and a target pitch-class sequence is computed and applied as an RMS-normalized update to the latent.
The base model is never retrained; the probe, containing approximately 125k parameters, is the only learned component.

Our contributions are twofold.
First, we demonstrate that frame-level pitch-class information is decodable from the Stable Audio Open VAE latent space with a micro-F1 score of 0.66, and that this decodability remains robust under diffusion noise levels encountered during the denoising process.
Second, we propose probe-guided latent steering for diffusion-based music generation and show that it significantly improves melodic coherence over unguided generation across a range of guidance strengths and target melodies.

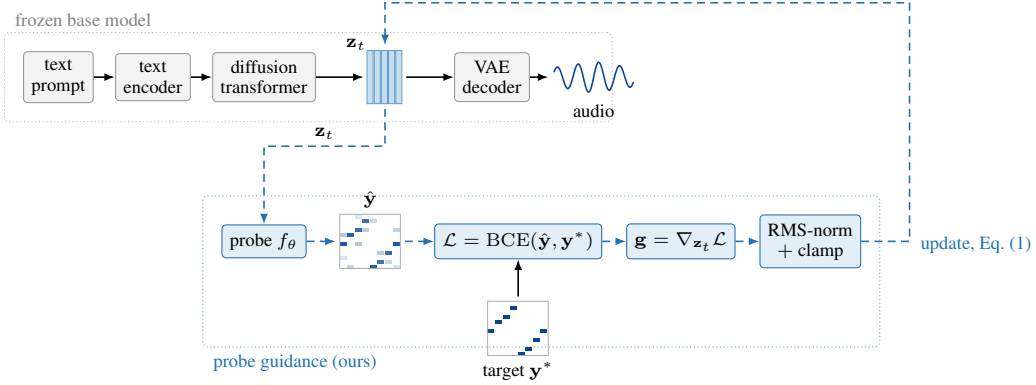
\begin{figure*}[t]
\centering
\begin{tikzpicture}[
  font=\scriptsize,
  frz/.style={draw=gry,fill=gry!12,rounded corners=1.5pt,inner sep=3pt,align=center},
  our/.style={draw=blB,fill=blA!15,rounded corners=1.5pt,inner sep=3pt,align=center},
  gen/.style={-{Latex[length=4pt]},semithick},
  grad/.style={-{Latex[length=4pt]},semithick,blB,densely dashed},
  lbl/.style={inner sep=1pt,font=\scriptsize}]

\node[frz]                    (p)    {text\\prompt};
\node[frz,right=8pt of p]     (clap) {text\\encoder};
\node[frz,right=8pt of clap]  (dit)  {diffusion\\transformer};
\coordinate (zc) at ($(dit.east)+(26pt,0)$);
\foreach \i in {0,1,2,3,4}{%
  \fill[blA!30,draw=blB!70,line width=.3pt]
    ($(zc)+(\i*2.7pt-6.75pt,-10pt)$) rectangle ++(2.3pt,20pt);}
\node[lbl,anchor=south east] at ($(zc)+(-6pt,9pt)$) {$\mathbf{z}_t$};
\node[frz,right=26pt of zc]   (dec)  {VAE\\decoder};
\coordinate (wv) at ($(dec.east)+(24pt,0)$);
\begin{scope}[shift={(wv)},x=1pt,y=1pt]
  \draw[blC,semithick] plot[domain=-15:15,samples=100,variable=\x]
    (\x,{6*sin(\x*46)*exp(-abs(\x)/30)});
\end{scope}
\node[lbl,anchor=north] at ($(wv)+(0,-9pt)$) {audio};

\draw[gen] (p)    -- (clap);
\draw[gen] (clap) -- (dit);
\draw[gen] (dit.east) -- ($(zc)+(-8pt,0)$);
\draw[gen] ($(zc)+(8pt,0)$) -- (dec.west);
\draw[gen] (dec.east) -- ($(wv)+(-17pt,0)$);

\node[our,below=46pt of dit.south,anchor=north] (probe)
  {probe $f_\theta$};
\coordinate (yh) at ($(probe.east)+(24pt,0)$);
\node[our,right=48pt of probe]  (loss) {$\mathcal{L}=\mathrm{BCE}(\hat{\mathbf{y}},\mathbf{y}^*)$};
\node[our,right=9pt of loss]    (grd)  {$\mathbf{g}=\nabla_{\mathbf{z}_t}\mathcal{L}$};
\node[our,right=9pt of grd]     (upd)  {RMS-norm\\$+$ clamp};

\begin{scope}[shift={($(yh)+(-11.5pt,-10.2pt)$)}]
  \foreach \c/\r/\o in {0/5/90,0/7/28,0/2/15,1/7/85,1/0/22,2/8/80,2/5/30,2/11/14,
                        3/10/78,3/8/26,4/0/86,4/10/20,5/1/76,5/3/28,6/3/80,6/1/22,
                        7/5/90,7/8/24,7/0/14}
    \fill[blC,opacity=\o/100] (\c*2.85pt,\r*1.7pt) rectangle ++(2.5pt,1.45pt);
  \draw[gry!80,line width=.3pt] (0,0) rectangle (22.8pt,20.4pt);
\end{scope}
\node[lbl,anchor=south] at ($(yh)+(0,11.5pt)$) {$\hat{\mathbf{y}}$};

\coordinate (yt) at ($(loss.south)+(0,-26pt)$);
\begin{scope}[shift={($(yt)+(-11.5pt,-10.2pt)$)}]
  \foreach \c/\r in {0/5,1/7,2/8,3/10,4/0,5/1,6/3,7/5}
    \fill[blC] (\c*2.85pt,\r*1.7pt) rectangle ++(2.5pt,1.45pt);
  \draw[gry!80,line width=.3pt] (0,0) rectangle (22.8pt,20.4pt);
\end{scope}
\node[lbl,anchor=north] at ($(yt)+(0,-11.5pt)$) {target $\mathbf{y}^*$};

\draw[grad] (probe.east)      -- ($(yh)+(-13pt,0)$);
\draw[grad] ($(yh)+(13pt,0)$) -- (loss.west);
\draw[grad] (loss) -- (grd);
\draw[grad] (grd)  -- (upd);
\draw[gen]  ($(yt)+(0,12pt)$) -- (loss.south);

\begin{pgfonlayer}{bg}
  \node[draw=gry!70,densely dotted,rounded corners=2pt,inner sep=7pt,
        fit=(p)(clap)(dit)(dec)(wv)] (boxA) {};
  \node[draw=blB!70,densely dotted,rounded corners=2pt,inner sep=7pt,
        fit=(probe)(loss)(grd)(upd)(yt)] (boxB) {};
\end{pgfonlayer}
\node[lbl,gry!85!black,anchor=south west] at ($(boxA.north west)+(3pt,1pt)$)
  {frozen base model};
\node[lbl,blB,anchor=north west] at ($(boxB.south west)+(3pt,-1.5pt)$)
  {probe guidance (ours)};

\coordinate (R) at ($(boxB.east)+(11pt,0)$);
\coordinate (T) at ($(boxA.north)+(0,11pt)$);
\coordinate (g1) at ($(zc)+(0,-26pt)$);
\coordinate (g2) at (probe.north |- g1);
\draw[grad] ($(zc)+(0,-12pt)$) -- (g1) -- (g2) -- (probe.north);
\node[lbl,anchor=south,black] at ($(g1)!0.5!(g2)+(0,1.5pt)$) {$\mathbf{z}_t$};
\draw[grad] (upd.east) -- (R |- upd) -- (R |- T) -- (zc |- T) -- ($(zc)+(0,12pt)$);
\node[lbl,blB,anchor=west] at ($(R)+(3pt,10pt)$) {update, Eq.~(\ref{eq:update})};
\end{tikzpicture}
\caption{Probe-guided latent steering at inference time. The base model is frozen (grey); the only learned component is the pitch-class probe $f_\theta$ (blue), a small $\sim$125k-parameter network trained offline on paired audio and MIDI and kept frozen at inference, which decodes a frame-level pitch-class map $\hat{\mathbf{y}}$ from the denoising latent $\mathbf{z}_t$. The binary cross-entropy loss against the user-supplied target $\mathbf{y}^*$ is backpropagated to $\mathbf{z}_t$ and applied as an RMS-normalized, clamped update. Both pitch-class maps are drawn schematically.}
\label{fig:overview}
\end{figure*}

\section{Related Work}
\label{sec:related}

Controllable music generation has been approached from several directions.
One family of methods modifies the generation model: Music ControlNet~\cite{wu2024music} trains additional conditioning modules for melody and dynamics control of diffusion models, and Mustango~\cite{melechovsky2024mustangocontrollabletexttomusicgeneration} conditions on musical attributes through text augmentation.
These approaches require retraining or architectural changes.
A second family operates at inference time: DITTO~\cite{Novack2024Ditto} and DITTO-2~\cite{Novack2024DITTO2DD} optimize differentiable losses during the denoising process of a music diffusion model but require hand-designed loss functions for each attribute and do not exploit latent-space geometry.
In the autoregressive setting, activation patching~\cite{facchiano2025activation}, probe weights repurposed as steering vectors~\cite{panda2025finegrained}, Recursive Feature Machines~\cite{zhao2025musicrfm}, and injected mean activation vectors~\cite{jiang2025composer} have all been used for attribute- or note-level control.
Our method bridges these directions: like DITTO it operates at inference time on a diffusion model, but like the autoregressive methods it derives its control signal from a learned probe.
The settings are complementary rather than directly competing: DITTO optimizes the sampler itself for each attribute and target and Music ControlNet trains conditioning modules, whereas we leave both the model and the sampler untouched and add only a frozen $\sim$125k-parameter probe as the source of the control gradient.

Probes have been widely used to assess what information neural representations encode~\cite{alain2016understanding}.
Wei et al.~\cite{wei2024musicgenerationmodelsencode} probe MusicGen for music-theoretic concepts, Koo et al.~\cite{Koo2024understanding} probe attention heads in music transformers, and Singh et al.~\cite{singh2025discovering} and Paek et al.~\cite{paek2025learning} train sparse autoencoders on music model activations to discover interpretable features.
Our work extends probing from a diagnostic tool to an active control mechanism: the probe both reveals what the latent space encodes and provides the gradient signal for steering.

Huang et al.~\cite{huang2025aligning} systematically study evaluation metrics for text-to-music systems and find that existing automated metrics correlate poorly with human judgment.
Following this insight, we use a task-specific melodic coherence metric rather than general-purpose audio quality scores.

\section{Method}
\label{sec:method}

Our approach consists of two stages: training a pitch-class probe on the VAE latent space (Section~\ref{ssec:probe}) and using the trained probe to guide the diffusion sampling process at inference time (Section~\ref{ssec:inference}); Figure~\ref{fig:overview} summarizes the resulting inference-time procedure.
We begin with a brief overview of the base model.

Stable Audio Open~\cite{evans2024open} is a text-to-music latent diffusion model.
It consists of three components: an Oobleck VAE that compresses waveforms into a continuous latent space with 64 channels, a text encoder (CLAP~\cite{wu2023large}) that produces conditioning embeddings from text prompts, and a diffusion transformer that generates latent sequences conditioned on both text and timing embeddings.
The VAE operates at a high compression ratio, mapping 5 seconds of 44.1\,kHz audio (220{,}500 samples) to a latent tensor of shape $64 \times T$ where $T \approx 215$ frames.
This compressed representation must retain enough musical information for the diffusion model to reconstruct coherent audio, making it an informative target for probing: if pitch information survives the compression, it can potentially be read out and used as a control signal.

\subsection{Pitch-Class Probe}
\label{ssec:probe}

We train a lightweight convolutional probe to predict frame-level pitch-class activations from the VAE latent representation of Stable Audio Open.
The input to the probe is a latent tensor $\mathbf{z} \in \mathbb{R}^{C \times T}$ with $C{=}64$ channels and $T$ time frames produced by the Oobleck VAE encoder.
The output is a multi-label prediction $\hat{\mathbf{y}} \in [0,1]^{T \times 12}$ indicating the presence of each of the 12 pitch classes (C, C$\sharp$, D, \ldots, B) at each frame.

The probe architecture is a two-layer one-dimensional convolutional network.
Each layer applies a 1D convolution with kernel size 5 and same-padding, followed by group normalization with 8 groups and a ReLU nonlinearity.
A final pointwise convolution (kernel size 1) projects the hidden representation to 12 output channels at each frame, and a sigmoid activation produces per-class probabilities.
The hidden dimension is 128, yielding approximately 125k trainable parameters.
This architecture captures short-range temporal dependencies while remaining small enough to provide a meaningful test of what the latent space encodes.

We train the probe on a portion of the MAESTRO dataset~\cite{hawthorne2018enabling}, which provides time-aligned audio and MIDI pairs of piano performances.
Audio clips of 5 seconds are encoded through the frozen VAE to produce latent representations, and pitch-class labels are extracted from the aligned MIDI files using the pretty\_midi library~\cite{raffel2014pretty}.
After splitting at the file level to prevent data leakage from clips originating in the same recording, we obtain 2{,}637 training clips and 417 validation clips.

Because the probe will be applied to partially denoised latents during inference, we augment the training data with additive Gaussian noise at scales $\sigma \in \{0, 0.1, 0.3, 0.5, 0.7\}$, sampled uniformly per batch element.
The noised latent is then RMS-normalized to preserve the original magnitude scale.
This augmentation ensures the probe generalizes to the range of noise levels encountered at different stages of the diffusion denoising process.

The probe is trained with binary cross-entropy loss using the AdamW optimizer with a learning rate of $10^{-3}$ and weight decay of $10^{-5}$, cosine learning rate annealing, and gradient clipping at norm 1.0, for a total of 50 epochs.

\subsection{Probe-Guided Inference}
\label{ssec:inference}

Given a text prompt and a target pitch-class sequence, we steer the standard diffusion sampling process of Stable Audio Open by using the trained probe as a differentiable loss function.

The user provides a list of MIDI note numbers and a tempo, which are converted to a frame-level pitch-class target $\mathbf{y}^* \in \{0,1\}^{T \times 12}$ by mapping each note to its pitch class and distributing note onsets across frames according to the specified tempo and note duration.
This representation abstracts away octave information, focusing the guidance signal on pitch-class content rather than absolute register.
We stress that the guidance target is this user-supplied symbolic pitch-class schedule, not audio or pitch content derived from the text prompt: the prompt determines style and timbre while the probe target determines pitch content, and the two are specified independently.

Stable Audio Open uses a denoising schedule of $N{=}50$ steps with classifier-free guidance at scale 4.0.
We apply probe guidance starting at step $\lfloor 0.4N \rfloor$ (after 40\% of denoising is complete) and continuing to the final step, intervening every 2 steps.
Guidance is not applied during the initial 40\% of denoising because the latent at early steps is dominated by noise and the probe's predictions are unreliable; allowing the diffusion model to establish coarse structure before steering begins produces better results than intervening from the start.

At each guidance step, the current latent $\mathbf{z}_t$ is cloned and passed through the probe in float32 precision to obtain pitch-class logits $\hat{\mathbf{y}} = f_\theta(\mathbf{z}_t)$.
The binary cross-entropy loss $\mathcal{L} = \mathrm{BCE}(\hat{\mathbf{y}}, \mathbf{y}^*)$ is computed between the probe predictions and the target, and the gradient $\mathbf{g} = \nabla_{\mathbf{z}_t} \mathcal{L}$ is obtained via backpropagation.
The latent is then updated as
\begin{equation}
\mathbf{z}_t \leftarrow \mathbf{z}_t - \mathrm{clamp}\!\left(\lambda \, \frac{\mathrm{rms}(\mathbf{z}_t)}{\mathrm{rms}(\mathbf{g})} \, \mathbf{g},\; \pm\alpha\,\mathrm{rms}(\mathbf{z}_t)\right),
\label{eq:update}
\end{equation}
where $\lambda$ is the guidance scale and $\alpha{=}0.05$ is the maximum update ratio.

\subsection{Theoretical Analysis}
\label{ssec:theory}

We motivate the update rule in Equation~\ref{eq:update} by connecting it to classifier guidance~\cite{dhariwal2021diffusion} and analyzing its stability properties.

In standard classifier guidance, the score function of a conditional diffusion model is decomposed as
\begin{equation}
\nabla_{\mathbf{z}} \log p(\mathbf{z}_t \mid y) = \nabla_{\mathbf{z}} \log p(\mathbf{z}_t) + \gamma \nabla_{\mathbf{z}} \log p(y \mid \mathbf{z}_t),
\label{eq:classifier_guidance}
\end{equation}
where $p(y \mid \mathbf{z}_t)$ is a noise-conditional classifier and $\gamma$ controls the guidance strength~\cite{ho2022cfg}.
Our setting differs from this formulation in one important respect: the probe $f_\theta$ is trained on VAE-encoded clean latents augmented with noise, but is not conditioned on the denoising step $t$.
The negative gradient of the BCE loss is therefore best understood as a step-agnostic approximation to $\nabla_{\mathbf{z}} \log p(\mathbf{y}^* \mid \mathbf{z}_t)$ rather than a strict implementation of classifier guidance.
Empirically, the noise augmentation during probe training (Section~\ref{ssec:probe}) and the stability properties analyzed below are sufficient to make this approximation effective in practice.

A key challenge in applying classifier guidance to latent diffusion models is that the gradient magnitude $\|\mathbf{g}\|$ varies by orders of magnitude across denoising steps due to the changing signal-to-noise ratio of $\mathbf{z}_t$.
Without normalization, a fixed step size $\gamma$ would produce negligible updates at some steps and destructive ones at others.
We address this through RMS normalization: for the unnormalized update direction $\boldsymbol{\delta} = \lambda \frac{\mathrm{rms}(\mathbf{z}_t)}{\mathrm{rms}(\mathbf{g})} \mathbf{g}$, the definition of $\mathrm{rms}(\cdot)$ and the scalar scaling give
\begin{equation}
\mathrm{rms}(\boldsymbol{\delta}) = \lambda \cdot \mathrm{rms}(\mathbf{z}_t),
\label{eq:rms_property}
\end{equation}
so the typical magnitude of each element in the update is a fixed fraction $\lambda$ of the typical magnitude of the latent, regardless of the gradient's absolute scale.
The update is therefore automatically adaptive: as $\mathrm{rms}(\mathbf{z}_t)$ decreases during denoising (the signal becomes cleaner), the update magnitude decreases proportionally, applying progressively gentler corrections in later steps where the latent carries more fine-grained structure.

The element-wise clamp at $\pm \alpha \cdot \mathrm{rms}(\mathbf{z}_t)$ provides a worst-case bound.
For any element $i$ of the updated latent $\mathbf{z}_t'$, we have
\begin{equation}
|z_{t,i}' - z_{t,i}| \leq \alpha \cdot \mathrm{rms}(\mathbf{z}_t),
\label{eq:clamp_bound}
\end{equation}
bounding the maximum per-element perturbation to a fraction $\alpha$ of the latent's overall scale.
With $\alpha{=}0.05$, no single guidance step can shift any latent element by more than 5\% of the RMS.
We found empirically that this bound prevents the mode collapse and audible artifacts observed at larger values of $\alpha$, while still permitting cumulative perturbation across the guidance window to meaningfully shift the generated pitch content.

\section{Experimental Setup}
\label{sec:experiments}

We evaluate probe-guided steering on a set of 27 trials constructed from 9 text prompts crossed with 3 target melodies.
The text prompts describe piano music with varying stylistic characteristics, ranging from classical and jazz to ambient and cinematic styles.
The three target melodies are: an ascending diatonic scale from F3 to F4 (8 notes), a descending scale from F4 to F3 (8 notes), and an alternating arpeggio F3--A$\flat$3--C4--F4--C4--A$\flat$3--F3 (7 notes).
All melodies are drawn from the F natural minor scale and use a tempo of 120 BPM with one note per beat, yielding target durations of 3.5 to 4 seconds within the 5-second generation window.
Generated audio is 5 seconds long at a sampling rate of 44.1\,kHz.

We compare four probe guidance scales ($\lambda{=}0.03$, $0.05$, $0.10$, and $0.20$) against an unguided baseline that uses the same prompts and random seed but applies no probe intervention.
All conditions share the same diffusion hyperparameters: 50 denoising steps, classifier-free guidance scale of 4.0, and a fixed random seed.

To quantify the degree of pitch-class control, we define a melodic coherence metric.
We first extract fundamental frequency contours from the generated audio using the pYIN algorithm~\cite{mauch2014pyin} as implemented in librosa~\cite{mcfee2025librosa}.
Detected pitches are converted to pitch classes and compared frame-by-frame against the target pitch-class schedule within regions where the target specifies an active note.
Melodic coherence is defined as the fraction of voiced frames in target-active regions where the detected pitch class matches the target; throughout this paper we use this term as shorthand for this frame-level pitch-class match rate, and it does not capture higher-order melodic properties such as interval structure or rhythmic alignment.
Trials in which fewer than 20\% of target-region frames are voiced are excluded to avoid unreliable estimates; in practice, no trials were excluded.

The absolute scale of this metric is not directly interpretable.
A naive $1/12 \approx 0.083$ rate is not the appropriate reference point, since a match requires the correct pitch class at the correct \emph{frame} of a time-varying target rather than a single class held over the clip.
Moreover, pYIN is a monophonic $f_0$ estimator while the generated piano audio is polyphonic, so a correctly realized target melody embedded in accompaniment is still scored as a mismatch whenever the tracker follows a concurrent voice, placing the practical ceiling well below 1.0.
We therefore measure the chance level empirically with a \emph{shuffled-target control}: every clip is additionally scored against all 11 non-trivial circular rotations of its own pitch-class target.
Rotation leaves the rhythm, the active-frame mask and hence the denominator of the metric unchanged, varying only pitch-class identity, so it isolates exactly the quantity the metric is meant to capture.
We report the mean over the 11 rotations and test each condition against its own shuffled control with a one-sided Wilcoxon signed-rank test.

To check that steering does not come at the expense of audio quality, we additionally report two objective metrics over the same 27 trials: CLAP~\cite{wu2023large} text--audio similarity as a measure of prompt adherence, and Fr\'echet Audio Distance (FAD) with OpenL3 embeddings against the MAESTRO recordings as a measure of audio quality.

Statistical significance is assessed using the Wilcoxon signed-rank test (one-sided) on paired per-trial coherence differences between each guided method and the baseline, computed across all 27 trials.

\section{Results}
\label{sec:results}

\subsection{Probe Decoding Performance}
\label{ssec:probe_results}

The convolutional probe achieves a micro-F1 score of 0.663 and a macro-F1 of 0.657 on held-out validation data, with per-class F1 scores ranging from 0.620 to 0.683 across all 12 pitch classes.
The positive-class recall, measured as the fraction of frames containing at least one active pitch class where the probe correctly identifies at least one, is 0.596.
These results confirm that the Stable Audio Open VAE latent space encodes pitch-class information in a form accessible to a small nonlinear probe, and the relatively uniform per-class performance suggests that no particular pitch class is privileged or suppressed, which is important for unbiased steering across arbitrary target melodies.

Table~\ref{tab:noise} examines how probe performance degrades as Gaussian noise is added to the latent representations, simulating the conditions encountered during diffusion denoising.
Performance decreases gradually from an F1 of 0.663 at $\sigma{=}0$ to 0.579 at $\sigma{=}1.0$, indicating that pitch-class information remains substantially accessible even at noise levels well beyond those used during training augmentation.
This robustness is essential for the guidance scheme, since the probe must produce a useful gradient signal when applied to partially denoised latents that still contain significant noise.

\begin{table}[t]
\centering
\caption{Probe decoding performance under additive Gaussian noise. The probe maintains usable accuracy across noise levels encountered during diffusion denoising.}
\label{tab:noise}
\vspace{0.5em}
\begin{tabular}{@{}lcc@{}}
\toprule
Noise $\sigma$ & F1 (micro) & Positive recall \\
\midrule
0.0 & 0.663 & 0.596 \\
0.1 & 0.662 & 0.595 \\
0.3 & 0.657 & 0.588 \\
0.5 & 0.642 & 0.569 \\
0.7 & 0.621 & 0.542 \\
1.0 & 0.579 & 0.493 \\
\bottomrule
\end{tabular}
\end{table}

\subsection{Melodic Coherence}
\label{ssec:coherence}

\begin{figure}[t]
\centering
\includegraphics[width=0.85\linewidth]{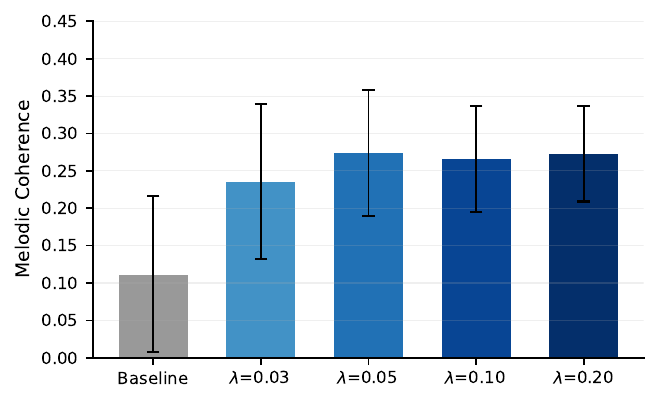}
\caption{Mean melodic coherence with standard deviation error bars across guidance scales. The baseline (gray) uses no probe intervention. All probe-guided conditions (blue) substantially improve pitch-class alignment.}
\label{fig:coherence}
\end{figure}

Table~\ref{tab:coherence} and Figure~\ref{fig:coherence} summarize the main results.
All four probe-guided conditions significantly outperform the unguided baseline, increasing mean melodic coherence from 0.112 to a range of 0.236 to 0.274 depending on the guidance scale.
The best-performing condition ($\lambda{=}0.05$) achieves a 2.4$\times$ improvement in mean coherence over the baseline, and all guided conditions yield highly significant improvements ($p < 5 \times 10^{-4}$).

\begin{table}[t]
\centering
\caption{Melodic coherence across probe guidance scales, with the shuffled-target control. $p_{\mathrm{base}}$ tests each condition against the unguided baseline; $p_{\mathrm{shuf}}$ tests it against its own shuffled control (one-sided Wilcoxon signed-rank test, $n{=}27$ paired trials).}
\label{tab:coherence}
\vspace{0.5em}
\setlength{\tabcolsep}{3pt}
\begin{tabular}{@{}lcccc@{}}
\toprule
Method & Mean $\pm$ Std & Shuffled & $p_{\mathrm{base}}$ & $p_{\mathrm{shuf}}$ \\
\midrule
Baseline        & 0.112 $\pm$ 0.104 & 0.081 & --- & 0.14 \\
Probe ($\lambda{=}$0.03) & 0.236 $\pm$ 0.104 & 0.069 & $4.6\!\times\!10^{-4}$ & $1.9\!\times\!10^{-7}$ \\
Probe ($\lambda{=}$0.05) & 0.274 $\pm$ 0.084 & 0.066 & $5.7\!\times\!10^{-6}$ & $7.5\!\times\!10^{-9}$ \\
Probe ($\lambda{=}$0.10) & 0.266 $\pm$ 0.071 & 0.067 & $8.0\!\times\!10^{-6}$ & $7.5\!\times\!10^{-9}$ \\
Probe ($\lambda{=}$0.20) & 0.273 $\pm$ 0.064 & 0.066 & $2.3\!\times\!10^{-6}$ & $7.5\!\times\!10^{-9}$ \\
\bottomrule
\end{tabular}
\end{table}

The shuffled-target control fixes the empirical chance level of the metric at 0.066--0.081, somewhat below the naive $1/12$.
It also sharpens the comparison in an important way: the unguided baseline is \emph{not} significantly above its own shuffled control ($p{=}0.14$), so its coherence of 0.112 is attributable to incidental pitch-class overlap between the prompt-driven output and the target rather than to any alignment with the requested melody.
Every probe-guided condition, in contrast, exceeds its shuffled control by a factor of 3.4 to 4.2 with $p \leq 1.9 \times 10^{-7}$.
The control value is itself slightly lower under guidance (0.066 vs.\ 0.081 for the baseline), consistent with guidance concentrating pitch content on the target classes and thereby leaving less to be matched by a rotated target.
Taken together, this establishes that the gains are specific to the requested pitch-class sequence and not an artifact of guidance making the audio easier for a monophonic tracker to follow.

Coherence improves rapidly from $\lambda{=}0.03$ to $\lambda{=}0.05$ and then plateaus, suggesting that a moderate guidance strength already exploits the available pitch-class signal in the latent space.
The standard deviation nonetheless decreases monotonically with guidance strength, from 0.104 at $\lambda{=}0.03$ to 0.064 at $\lambda{=}0.20$, indicating that stronger guidance produces more consistent steering across diverse prompts and melody types even once the mean improvement has leveled off.

Table~\ref{tab:melody} breaks down the results by melody type at the best operating point ($\lambda{=}0.05$).
All three target patterns show substantial improvements over the baseline, confirming that the steering mechanism generalizes across different pitch-class trajectories.
The alternating arpeggio achieves the highest guided coherence, possibly because its wider pitch-class intervals create a more distinctive target signal for the probe gradient.

\begin{table}[t]
\centering
\caption{Per-melody coherence breakdown at $\lambda{=}0.05$ ($n{=}9$ prompts per melody type). All melody types show substantial improvement.}
\label{tab:melody}
\vspace{0.5em}
\begin{tabular}{@{}lcc@{}}
\toprule
Melody & Baseline & Probe ($\lambda{=}$0.05) \\
\midrule
Ascending  & 0.095 $\pm$ 0.081 & 0.268 $\pm$ 0.074 \\
Descending & 0.144 $\pm$ 0.108 & 0.234 $\pm$ 0.064 \\
Alternating & 0.096 $\pm$ 0.112 & 0.320 $\pm$ 0.090 \\
\bottomrule
\end{tabular}
\end{table}

\subsection{Audio Quality}
\label{ssec:quality}

Perturbing the denoising trajectory risks degrading the audio itself, so we verify that the coherence gains are not bought at the cost of quality.
CLAP text--audio similarity is essentially unchanged between the unguided baseline (0.274) and probe-guided generation (0.279), with no significant difference under a Wilcoxon signed-rank test ($p{=}0.75$); even at stronger guidance no significant degradation is observed ($p{=}0.20$).
FAD-OpenL3 against MAESTRO is likewise not worse under guidance (260.4 guided vs.\ 267.7 unguided).
Neither metric indicates a quality penalty, supporting the interpretation that the probe gradient shifts pitch content within the region of latent space the model already considers plausible rather than pushing the latent off the learned manifold.

\section{Discussion}
\label{sec:discussion}

The results demonstrate that the VAE latent space of a diffusion-based music model encodes pitch-class information that is not only decodable but also actionable for generation steering.
The probe's moderate F1 of 0.66 might suggest limited utility as a control signal, yet the downstream coherence improvements are large and statistically robust.
Steering does not require perfect frame-level predictions, only that the gradient of the probe loss point toward higher likelihood of the target pitch content; an imperfect probe still provides a useful directional signal because the gradient aggregates information across all pitch classes and time frames simultaneously.

The adaptive behavior predicted by the analysis in Section~\ref{ssec:theory} is confirmed empirically: RMS normalization produces naturally gentler corrections in later denoising steps where the latent carries more fine-grained structure, without requiring a manually tuned step-size schedule.
The saturation of coherence beyond $\lambda{=}0.05$, combined with decreasing variance at higher scales, suggests that stronger guidance primarily reduces outlier trials rather than improving the typical case, a desirable property for practical use.

All experiments train the probe on the MAESTRO piano dataset while evaluating on diverse piano-related text prompts, demonstrating a domain transfer that is likely facilitated by pitch-class structure being a low-level musical property that generalizes across piano styles.
Our evaluation is limited to a single base model (Stable Audio Open), pitch-class control without full melodic specification including rhythm and register, and simple diatonic test melodies.
It is also limited to piano-centric prompts: the probe is trained on solo piano, and how reliably it localizes pitch content in denser mixed-instrument textures, where several timbres overlap in the same latent channels, remains an important open question.
We leave generalization to other diffusion models, richer musical targets, mixed-instrument material, and broader timbral evaluation to future work.

\section{Conclusion}
\label{sec:conclusion}

We have presented a lightweight approach to pitch-class steering in diffusion-based music generation.
By training a small convolutional probe on the VAE latent space of Stable Audio Open and repurposing it as a differentiable loss at inference time, we achieve significant and consistent improvements in melodic coherence without retraining the base model.
The method requires only approximately 125k additional parameters, adds minimal computational overhead, and accommodates arbitrary pitch-class targets at inference time.
More broadly, our results suggest that latent diffusion models for music encode musically meaningful structure in their intermediate representations, and that lightweight probes offer a practical means of both analyzing and exploiting it.
The probe-as-loss paradigm is not specific to pitch: the same framework could in principle be applied to any musical attribute decodable from the VAE latent space, such as chord quality, rhythmic density, or timbre, by training an appropriate probe and using its gradient as the steering signal.

\bibliographystyle{IEEEbib}
\bibliography{refs}

\end{document}